\documentclass[aps,pra,showpacs,twocolumn,superscriptaddress]{revtex4-1}
\usepackage{graphicx,color}
\usepackage{dcolumn}
\usepackage{bm}
\usepackage{color}
\usepackage{amsmath,amsthm,amsfonts,amssymb,bm}
\usepackage{times}
\usepackage{physics}
\usepackage[normalem]{ulem}

\usepackage[hidelinks]{hyperref}
\usepackage[table,xcdraw]{xcolor}
\hypersetup{
    colorlinks,
    linkcolor={blue!80!black},
    citecolor={blue!80!black},
    urlcolor={blue!80!black}
}

\def\bra#1{\mathinner{\langle{#1}|}}
\def\ket#1{\mathinner{|{#1}\rangle}}

\begin{document}

\title{Continuous dynamical decoupling and decoherence-free subspaces for qubits \\with tunable interaction}

\author{\.{I}. Yal\c{c}\i nkaya}
\affiliation{Departments of Physics, Faculty of Nuclear Sciences and Physical Engineering, Czech Technical University in Prague, B\v{r}ehov\'a 7, 11519, Praha 1-Star\'e M\v{e}sto, Czech Republic}
\author{B. \c{C}akmak}
\affiliation{Department of Physics, Ko\c{c} University, 34450, \.{I}stanbul, Sar\i yer, Turkey}
\affiliation{College of Engineering and Natural Sciences, Bah\c{c}e\c{s}ehir University, 34353, Be\c{s}ikta\c{s}, \.{I}stanbul, Turkey}
\author{G. Karpat}
\affiliation{Department of Physics, Faculty of Arts and Sciences, Izmir University of Economics, 35330, \.{I}zmir, Turkey}
\author{F. F. Fanchini}
 \email{fanchini@fc.unesp.br}
\affiliation{Faculdade de Ci\^encias, UNESP - Universidade Estadual Paulista, Bauru, SP, 17033-360, Brazil}
\date{\today}

\begin{abstract}
Protecting quantum states from the decohering effects of the environment is of great importance for the development of quantum computation devices and quantum simulators. Here, we introduce a continuous dynamical decoupling protocol that enables us to protect the entangling gate operation between two qubits from the environmental noise. We present a simple model that involves two qubits which interact with each other with a strength that depends on their mutual distance and generates the entanglement among them, as well as in contact with an environment. The nature of the environment, that is, whether it acts as an individual or common bath to the qubits, is also controlled by the effective distance of qubits. Our results indicate that the introduced continuous dynamical decoupling scheme works well in protecting the entangling operation. Furthermore, under certain circumstances, the dynamics of the qubits naturally led them into a decoherence--free subspace which can be used complimentary to the continuous dynamical decoupling. 

\end{abstract}
\maketitle

\section{Introduction}

The interaction between a quantum system and its environment is primarily responsible for the loss of essential quantum features, such as quantum coherence and entanglement, which is widely known as decoherence \cite{zurek,schlosshauer}. However, it is these fragile features that make quantum systems advantageous in many different information processing tasks which significantly outperform their classical counterparts \cite{montanaro}. Therefore, it is compulsory to protect the quantum systems from the decohering effects of their environment to employ them in quantum computing protocols. 

There are several well-known strategies for protecting a quantum system \cite{steane,shor96,lidar98,lidar03}. One of the most effective ways is the dynamical decoupling (DD) protocols which have been studied both theoretically \cite{viola98,viola99,zanardi,byrd,facchi05,khodjasteh08,khodjasteh09feb,khodjasteh09sep,khodjasteh10,lidar2012review} and experimentally \cite{biercuk,du,damodarakurup,delange,souza,naydenov,vandersar}. The main idea of DD is to preserve the quantum features of the subject system by applying external pulses to eliminate the effect of the environment. Mathematically, this corresponds to introducing an external control Hamiltonian for the subject system, which cancels out the undesired dynamics arising from the system--environment coupling. Instead of applying a sequence of pulses, one can also protect the quantumness of a system by applying continuous external fields which is known in the literature as continuous dynamical decoupling (CDD) \cite{romero,chen,clausen,xu,fanchini07feb,fanchini07dec,fanchini2015,rabl,chaudhry,cai,laraoui}. Experimentally, application of two qubit gates is fairly more natural when the system is protected by CDD \cite{bermudez2011,bermudez2012,timoney} and also, CDD plays an important role in reducing the error induced by environmental perturbations in nitrogen vacancy centers in diamonds which are powerful candidates for the applications in the field of quantum information technologies \cite{cai,dohterty,albrecht,golter}. 

Besides the CDD protective scheme, one can also explore other aspects that may emerge during the dynamics. For certain system--environment interactions, there are some parts of the system Hilbert space that are unaffected by the decohering dynamics, therefore preserving the quantum information encoded in them. Such parts of the Hilbert space are called decoherence--free subspaces (DFSs), and they also constitute a very important place among the strategies to preserve quantum information \cite{lidar98,lidar03}. Encoding the desired information in these parts of the Hilbert space, of course, presents a very natural opportunity to transfer or process it without getting affected by the decoherence \cite{bacon,kempe}, and this approach has also found many experimental applications \cite{kielpinski,viola2001,boulant2002,mohseni2003,bourennane12004,altepeter2004,langer2004,monz2009,wang2017}.

In this work, we propose a model involving two qubits interacting with each other as well as with a bosonic environment. We assume that it is possible to control the exchange interaction between the qubits by changing their mutual distance. By varying this effective parameter, we both tune the interaction strength between the qubits and determine the way they couple to the environment. To be more precise, when the qubits are well separated, the interaction between them vanishes and they can be considered as if they are coupled to independent environments due to the large relative separation between them in the position space. In the opposite limit, in which qubits are close to each other, the interaction strength reaches its maximum and they are assumed to be coupled to a common environment. Our model is built in such a way that the transition between these limits is gradual, and both inter-qubit interaction strength and the environment coupling change simultaneously, since they are both controlled by the same parameter. In the absence of an environment, the interaction between the qubits is chosen so that after a certain time qubits become maximally entangled. However, due to the environmental noise, which is present in realistic experimental situations, it is not possible to reach this ideal entangled state. Such degrading effects of environment on the entanglement qubits are simulated in this work by amplitude damping and dephasing channels. We then present a CDD scheme designed specifically for our model to eliminate these harmful effects of the environment. Moreover, in some certain cases, for example depending on the final positions of qubits, we show that it is possible end up with a state in a DFS, which is indifferent to external noise even when the protection is switched off.

This paper is organized as follows. In Sect.~\ref{model} we present our CDD strategy and introduce the decoherence model that we consider together with its solution. Our results are presented in Sect.~\ref{results} and in Sect.~\ref{conclusion} we conclude.

\section{The model} \label{model}

In the interaction picture, the total Hamiltonian of the system under consideration is given in the form
\begin{equation}\label{dd}
  H(t)=H_{\rm {gate}}+H_{\rm {env}}+U^{\dagger}_cH_{\rm {int}}U_c.
\end{equation}
Here, $H_{\rm {gate}}$ defines the interaction between the two qubits which performs the entangling $\sqrt{\textsc{SWAP}}$ gate operation and $H_{\rm {env}}$ denotes the Hamiltonian of the environment. 
The last term includes the system--environment interaction $H_{\rm {int}}$, and $U_c$ is the time evolution operator corresponding to the CDD control Hamiltonian $H_c$. The interaction picture transformation leaves $H_{\rm {env}}$ intact since $U_c$ only affects the Hilbert space of the protected system. This is also true for $H_{\rm {gate}}$, which will be evident shortly after.

The tunable Heisenberg exchange interaction between the qubits can be written as
\begin{equation}
H_{\rm gate}=J(t){\bm \sigma}^{(1)}.{\bm \sigma}^{(2)},
\label{H0}
\end{equation}
for $\hbar\!=\!1$ and ${\bm \sigma}^{(s)}\!=\!{\bf
\hat{x}}\sigma_{x}^{(s)}+{\bf \hat{y}}\sigma_{y}^{(s)}+{\bf
\hat{z}}\sigma_{z}^{(s)}$, where $\sigma_{i}^{(s)}$'s ($s\!=\!1,2$) are the Pauli matrices acting on qubit $s$. In our model, we interpret the time dependence of $J$ as being adjusted due to the alteration of the qubit separation. Without loss of generality, we assume that coupling between qubits is in the form of a Gaussian function of time given as $J(t)\!=\!A\exp[-B(t-\tau/2)^2]$ where $\tau$ is the total interaction time.  Therefore, by using the known integral $\int_{-\infty}^{+\infty}J(t){\hbox{d}}t\!=\!A\sqrt{\pi/B}$, we can safely assert the constraint $B\!=\!(8A)^2/\pi$ as long as $\tau/2$ is greater than three-sigma interval of $J$. We make use of two different $J$ profiles specified by their peak heights $A$ corresponding two different physical situations which will be explained later. From this point on, we will refer to the time $t$ as being scaled by $\tau$.

There are many advantages of considering a tunable exchange interaction between the qubits, for both the theoretical and experimental aspects of our model. First of all, it can be realized in double quantum dot systems by controlling the intermediate tunnel barrier \cite{loss}. Among others such as voltage-controlled exchange \cite{petta2005}, one of the possible ways of achieving this control is to adjust the separation between the dots \cite{burkard,usaj}, which actually constitutes the motivation behind our aforementioned interpretation. It has also been shown experimentally these kinds of couplings can be realized with neutral atoms trapped in an optical lattice \cite{anderlini2007}, where initially separated and fully isolated Rubidium atoms are then gradually merged to occupy the same physical location and are allowed to interact. By this way, it is possible to create a tunable exchange constant $J(t)$, and hence, to realize SWAP and $\sqrt{\hbox{SWAP}}$ operations. Besides, Eq. (\ref{H0}) remains invariant under global rotations due to its scalar product form. As a result, the external control fields that are applied to protect the gate do not affect the gate operation itself, which in principle makes the concatenation of the protection mechanism and the entangling operation conceivable. Lastly, it is worthy mentioning that the tunable Heisenberg interaction alone is sufficient for universal quantum computation, i.e., it suffices to implement any quantum circuit without the necessity of having access to single-qubit operations \cite{divincenzo}, which is essential for quantum information processing.

\begin{figure}
\centering
\includegraphics[width=0.45\textwidth]{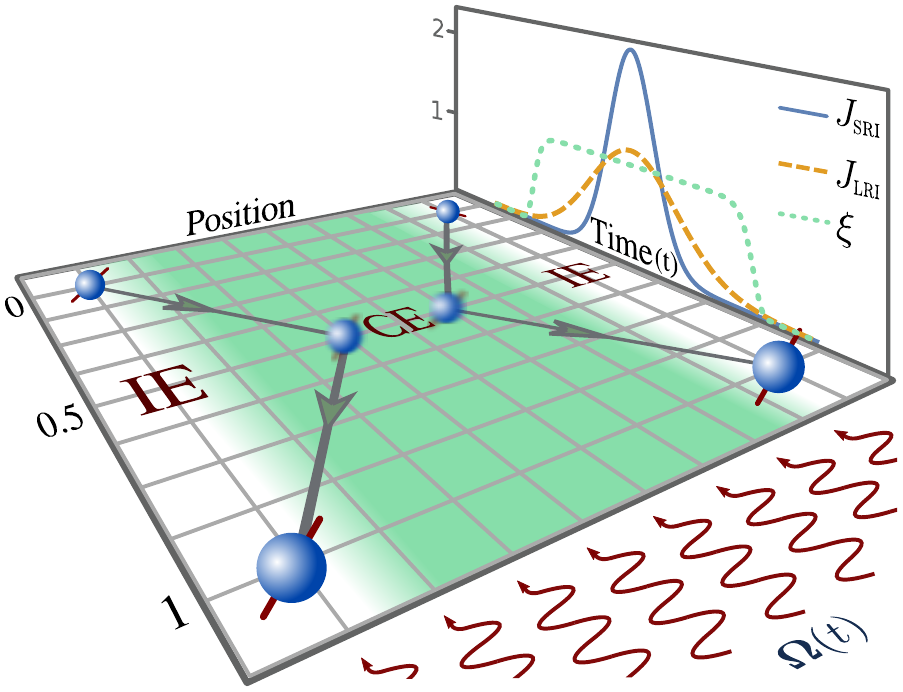}
\caption{Schematic representation of the presented model. At $t=0$, the non entangled qubits are well separated so that we assume the Heisenberg interaction among them is zero ($J=0$) and also, they are considered as interacting with two independent environments (IEs). As qubits get closer for $t\in[0,0.5]$, the interaction strength $J$ increases and the IE behavior gradually turns to a common environment (CE) behavior, which is characterized by the parameter $\xi$ as defined in Eq. \ref{eq:gradChEnv}. The same mechanism is reversed when $t\in[0.5,1]$. We choose $J(t)$ in two different ways where $J_\textsc{sri}$ ($J_\textsc{lri}$) corresponds to a short (long) range interaction effectively arising and vanishing in the CE (IE) region. In both cases, interactions are adjusted to lead the qubit pair evolve into a maximally entangled state in absence of environmental effects. During the whole interaction time $t \in [0,1]$, the system is kept in an external field $\bm \Omega(t)$ as given in Eq. \ref{Omega} to protect this entanglement operation from the environmental noise with an appropriate CDD procedure introduced in the text.}
\label{fig:1}
\end{figure}

The environmental Hamiltonian of both qubits is represented by a single thermal bath of harmonic oscillators. However, we assume that if the qubits are well-separated, their interaction with the environment can effectively be regarded as if they are in contact with two independent environments, e.g., as for the case of tunable charge qubits \cite{makhlin2001}. Therefore, we define the environmental Hamiltonian $H_{\rm env}$ in two different ways, depending on whether the qubits are coupled with a common environment (\textsc{CE}) or with independent environments (\textsc{IE}s). In the case of a \textsc{CE}, we have $H_{\rm env}\!=\!\sum _{k} \omega_{k} {a_{k}}^{\dagger}a_{k}$, where $\omega _{k}$ is the frequency of the $k$th normal mode of  the environment, and $a_{k}$ and ${a_{k}}^\dagger$ are the annihilation and creation operators, respectively. In the case of \textsc{IE}s, it can be written as $H_{\rm env}\!=\!\sum_{s=1}^{2}\sum_{k} \omega_{k}^{(s)}{a_{k}^{(s)}}^{\dagger}a_{k}^{(s)}$, where $\omega _{k}^{(s)}$ is the frequency of the $k$-th normal mode of the $s$th qubit environment. We assume that IEs are identical, i.e., the frequencies $\omega_{k}^{(s)}$ are the same and $\omega_k$ for both.

The qubit pair interacts with this bosonic environment according to the interaction Hamiltonian which is given by
\begin{equation}
H_{\rm int}={\bf B}^{(1)}\cdot {\bm \sigma}^{(1)}+{\bf B}^{(2)}\cdot {\bm \sigma}^{(2)},\label{eq:hint}
\end{equation}
where ${\bf B}^{(s)}=B_{x}^{(s)}{\bf\hat{x}}+B_{y}^{(s)}{\bf\hat{y}}+B_{z}^{(s)}{\bf\hat{z}}$ are Hermitian operators that act on the environmental Hilbert space. Accordingly, we take $B_{m}^{(s)}\!=\!\sum_{k}\left(\lambda_{m}g_{k}^{\ast}a_{k}^{(s)}+ \lambda_{m}^{\ast}g_{k}a_{k}^{(s)\dagger}\right)$, where $g_{k}$ are coupling constants. In the cases of CE or IE, Eq.~\eqref{eq:hint} reads as
\begin{eqnarray}
  H_{\rm int}^{\rm CE} &=& \left({\bm \sigma}^{(1)} + {\bm \sigma}^{(2)}\right)\cdot\left({\bm \lambda}B + {\bm\lambda }^*B^\dagger\right), \label{eq:hic}\\
  H_{\rm int}^{\rm IE} &=& {\bm \sigma}^{(1)}\cdot\left({\bm \lambda^{(1)}}B^{(1)} + {\bm\lambda^{(1)}}^\ast {B^{(1)}}^\dagger\right) \label{eq:hii} \\
  &+& {\bm \sigma}^{(2)}\cdot\left({\bm \lambda^{(2)}}B^{(2)} + {\bm\lambda^{(2)}}^\ast {B^{(2)}}^\dagger\right). \nonumber
\end{eqnarray}
For the CE case, ${\bm B}^{(s)}\!=\!{\bm \lambda}B+{\bm \lambda}^{\ast}B^{\dagger}$ where ${\bm \lambda}$ is an arbitrary complex three-dimensional vector and $B$ is a scalar operator that acts on the environmental Hilbert space. Similarly, for the IE case,  $B^{(s)}$ acts for the $s$th qubit and ${\bm \lambda^{(s)}}$ is the respective complex vector. As we have explained earlier, we will consider a smooth transition from IE to CE and then, CE to IE as qubits get closer and get departed from each other during the interaction time $\tau$, respectively. Therefore, we combine Eqs.~\eqref{eq:hic} and \eqref{eq:hii} into one effective Hamiltonian modeling the environmental interaction as
\begin{equation}
  H_{\rm int}= [1-\xi(t)]H_{\rm int}^{\rm IE}+ \xi(t) H_{\rm int}^{\rm CE},
  \label{eq:gradChEnv}
\end{equation}
where $\xi(t)\!=\!\exp[-\left[c(t-0.5)\right]^{2d}]$  defines the transition between the two interpretations of system--environment coupling as the qubits move with respect to each other. We will arbitrarily fix the parameters of $\xi(t)$ as $c\!=\!2.86$ and $d\!=\!9$ so that it yields a transition profile of smoothly interchanging low and high plateaus as we desire (see Fig. \ref{fig:1}). After having $H_{\rm int}$ fixed, now we define the $\sqrt{\textsc{SWAP}}$ in two different ways as mentioned earlier. (i) A short-ranged interaction (SRI) scheme denoted by $J_\textsc{sri}$ with  $A=2.2$, where whole interaction effectively arises and vanishes in the CE. (ii) A long-range interaction (LRI) scheme with parameters $A=1.15$, denoted by $J_\textsc{lri}$ where whole interaction effectively arises and vanishes in the IE. We note that the means of both $J(t)$ and $\xi(t)$ coincide, and they define the middle of the whole interaction at $t=\!0.5$. All parameters here were chosen especially to study the integration between CDD and DFS in different experimental situations.

Before moving on to the introduction of the control Hamiltonian, we need to introduce a necessary condition that must be satisfied by the control Hamiltonian in order for it to completely eliminate the effects of environment which can be mathematically expressed as \cite{viola98,viola99,zanardi,byrd,facchi05,khodjasteh08,khodjasteh09feb,khodjasteh09sep,khodjasteh10,lidar2012review}
\begin{equation}
\int ^{\tau}_{0} U^{\dagger}_{c}(t)H_{\rm int}U_{c}(t) \hbox{d}t=0,\label{condition}
\end{equation}
where  $\tau\!=\!2\pi /\omega $. In fact, Eq. \eqref{condition} is derived from a Magnus expansion of the total Hamiltonian given by Eq. \ref{dd} in the limit $\tau\rightarrow 0$, where, in general, only the first term in this expansion survives. Therefore, although in the ideal case of $\tau\rightarrow 0$ this approach works fine, in this work, we will consider the more realistic case of finite $\tau$ and expect Eq. (\ref{condition}) to also guide us in this case.

We propose the form of the control Hamiltonian that protects the entangling gate operation realized by Eq.~\ref{H0} to be
\begin{eqnarray}
H_{c}(t)={\bm \Omega}(t)\cdot\left( {\bm \sigma}^{(1)}+{\bm \sigma}^{(2)}\right), \label{hc}
\end{eqnarray}
where ${\bm \Omega}(t)$ is the external field configuration that needs to be applied. It has been shown that for Eq. (\ref{condition}) to be satisfied by the evolution operator corresponding to Eq. (\ref{hc}), the following equality has to be satisfied \cite{fanchini07dec}
\begin{eqnarray}
U_{c}(t)=U^{(2)}(t)U^{(1)}(t)=U^{(1)}(t)U^{(2)}(t),\label{Uc}
\end{eqnarray}
due to the fact that ${\bm \sigma}^{(1)}$ and ${\bm \sigma}^{(2)}$ commute, where
\begin{eqnarray}
U^{(s)}(t)=\exp\left(-i\omega t n_{x}\sigma_{x}^{(s)}\right)
\exp\left(-i\omega t n_{z}\sigma_{z}^{(s)}\right),\label{Uk}
\end{eqnarray}
for $s=1,2$. As a consequence, we obtain the external field configuration as
\begin{eqnarray}
{\bm \Omega}(t)={\bf \hat{x}}n_{x}\omega +n_{z}\omega \left[{\bf
\hat{z}}  \cos\left( n_{x}\omega t \right)-{\bf \hat{y}}
\sin\left( n_{x}\omega t \right) \right], \label{Omega}
\end{eqnarray}
with conditions imposed by Eqs.~\eqref{Uc} and \eqref{Uk}. Here, $\omega\!=\!2\pi/\tau$, $n_{x}$ and $n_{z}\!\ne\! n_{x}$ are nonzero  integers. This field configuration is composed of a combination of a static field along the $x$-axis and a rotating field around on the $yz$ plane. In this form, $H_c$ is capable of protecting our two-qubit system against both amplitude damping and dephasing errors. It is possible to consider a simpler field configuration by setting $n_z\!=\!0$ and still possible to provide protection solely against dephasing errors with a static field in the $x$-direction. It is worthy mentioning here that both dephasing and amplitude damping noises solely arise due to the interaction between two qubits and the environment. We assume that all other possible sources of noise, e.g., technical limitations caused by the driving of qubits, are negligible in our work. In other words, we are aware of the fact that amplitude fluctuations in the driving field can also introduce a noise to the system qubits; however, we assume that such fluctuations are unimportant for protection scheme considered in this work.

The reduced dynamics of the two qubits under consideration, which is dictated by the total Hamiltonian Eq.~(\ref{dd}), is governed by a Redfield master equation. The derivation and the solution of this master equation is elaborated in ``Appendix'', and also, an even more detailed explanation can be found in Refs. \cite{fanchini07dec,fanchini2015}.

\section{Results} \label{results}

In the following sections, we consider two different scenarios to introduce environmental noise on the entangling dynamics of our pair of qubits. First, we assume that dephasing is the only source of noise acting on the system and we investigate how well our system is protected against it for different values of the protective field strength. We also adjust the final positions of the qubits to leave them in contact with a CE after $J$ vanishes, so that we are able to make the two-qubit state to stay in a \textsc{DFS}. Second, we also let the amplitude damping to act on qubits together with dephasing and we examine how our protection scheme works in this case. Moreover, we show how the dynamics of entanglement is affected when either one of the error mechanisms is left as a residual error, i.e., no protection is provided against it. 

\begin{figure*}[t]
\centering
\includegraphics[width=.8\textwidth]{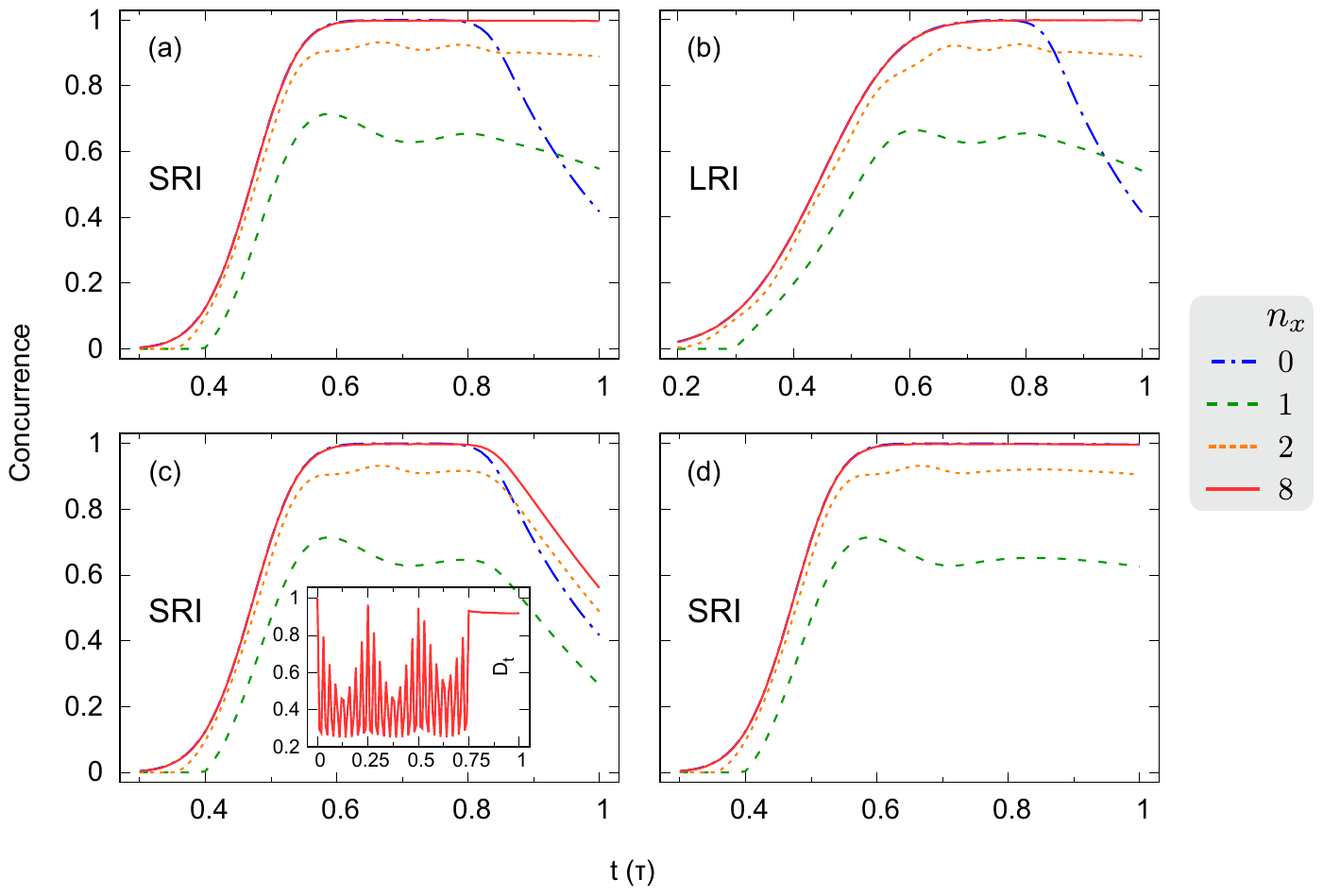}
\caption{Concurrence of the two-qubit state $\rho_t$ along the interaction time $\tau$ for increasing values of the protection $n_x$. For all cases, the system is only subjected to dephasing and hence, the $n_z$ component of the protective field is set to zero. The abbreviations \textsc{SRI} and \textsc{LRI} correspond to short- and long-range interaction schemes, respectively. In (a) and (b), the protection is maintained during the time evolution. In (c) and (d), the protection is shut down ($n_x=0$) after $t=0.75$, whereas in (d) the qubits are also halted after this time. Inset of (c) represents the DFS occupancy $D_t$ at time $t$, where $1$ ($0$) implies that the state $\rho_t$ is completely in (out of) the DFS. No further change in concurrence is observed for larger values of $n_x$. In (d), the curves coincide for $n_x=0$ and $n_x=8$.}
\label{dephas}
\end{figure*}

The initial state of our two qubit system is $\rho_0\!=\!|\uparrow\downarrow\rangle\langle\uparrow\downarrow|$. In the absence of any external noise and protection, $\sqrt{\textsc{SWAP}}$ gate --for both \textsc{SRI} and \textsc{LRI}-- yields the maximally entangled state $\rho_\tau\!=\![a\ket{\uparrow\downarrow}+a^*\ket{\downarrow\uparrow}][a^*\bra{\uparrow\downarrow}+a\bra{\downarrow\uparrow}]$ with $a\!=\!(1+i)/2$ at the end of the dynamics. Therefore, we examine the concurrence of this state with respect to time as a figure of merit when both the noise and the protection are introduced. The temperature of the environment(s) is chosen to be $T\!=\!0.2$ K, and we fix the relevant time scale in the dynamics to $\tau\!=\!10^{-9}$s. The environmental spectral density is chosen to be ohmic. Recall that, in all the different scenarios that will be considered in the following sections, initially the qubits are well separated, not interacting and in contact with the independent environments.

\subsection{Dephasing}

In this section, we  assume that the errors introduced on the system are caused only by dephasing. Therefore, we do not need to provide any protection against amplitude damping errors. We modify our protective field for this case by simply setting $n_z\!=\!0$. In Fig.~\ref{dephas}, we represent how the entanglement between the two qubits changes in time for different values of the protection strength, namely for $n_x\!=\!0,1,2$ and $8$, where $n_x\!=\!0$ means no protection at all.

While in Fig.~\ref{dephas}(a), (c) and (d) the coupling between the qubits effectively arises and vanishes when they are interacting with a common environment, Fig.~\ref{dephas}(b) presents the same situation but this time for independent environments (see Fig.~\ref{fig:1}). In other words, the former are the cases of a \textsc{SRI} and the latter is the case of a \textsc{LRI} among the qubits, as mentioned in Sec.~\ref{model}. First thing to notice in all cases is that as the strength of the external protection field increases, our CDD scheme works better and $n_x=8$ proves to be sufficient to fully protect the entangling gate operation. Even with $n_x\!=\!2$, it is possible to achieve a concurrence value of $\approx 0.9$. Thus, the present CDD protocol, which had proven to work for static inter-qubit coupling \cite{fanchini07dec,fanchini2015}, also performs completely well for the tunable case in question. Another point which also applies for all cases is that in the absence of any protection, qubits are getting highly entangled during a short time period before they start to get gradually non entangled because of the noise induced by the interaction with IEs. The concurrence for $n_x\!=\!0$ even exceeds that of the $n_x\!=\!2$ and reaches to the level of $n_x\!=\!8$. Nevertheless, the protection is still required if we take into account the whole interaction time $\tau$, i.e., setting the protection strength to at least $n_x\!=\!8$ is inevitable to obtain highly entangled states at $t\!=\!\tau$ for Fig.~\ref{dephas}(a) and (b). For the higher values of $n_x\!>\!8$, no further change in concurrence takes place.

We now turn our attention to the more interesting point of utilizing \textsc{DFS}s for protection after the qubits are entangled. First of all, in general, a \textsc{DFS} is only possible when the qubits are interacting with a common dephasing bath. In particular, the Hilbert space of two qubits, \textsc{DFS} is spanned by the following set of basis states $\mathcal{D}\!=\!\{\ket{\uparrow\downarrow}, \ket{\downarrow\uparrow}\}$. We have stated that, in the ideal case of isolated qubits, the initial state we consider ends up in a maximally entangled state in this subspace. Therefore, we want to see whether it is also possible to make use of this naturally occurring phenomenon for protection. At this point, we emphasize that there are two crucial requirements that need to be satisfied in order to maintain the existence of the two-qubit state in DFS. First of all, we must keep the qubits interacting with a CE after the inter-qubit interaction does its duty of entangling the qubits and vanishes. We can manage to do this in the SRI setting where the interaction begins and ends in the CE regime. The second condition is to turn the CDD field off since it drastically drives the two qubit state into and out of the DFS in time. To quantify how well the state $\rho_t$ is contained in the DFS, we can define 
\begin{equation}
D_t=\frac{1}{\xi_t}\sum_{i,j\in \mathcal{D}} \left|\mel{i}{\rho_t}{j}\right|.
\end{equation}
Here, $\xi_t^{-1}$ is the normalization factor which is obtained by summing over the absolute values of all elements in $\rho_t$. Thus, while $D_t\!=\!1$ implies a complete confinement of the $\rho_t$ in the DFS, the contrary case of $D_t\!=\!0$ indicates that the state is completely out of the DFS. In Fig.~\ref{dephas}(c), we show the dynamics of entanglement when the protection is turned off after $t\!=\!0.75$ for a SRI scenario by knowing in advance that the $\sqrt{\textsc{SWAP}}$ gate operation is completed and the qubits are maximally entangled before $t\!=\!0.75$. It can be seen from the inset that $D_t$ approximately reaches its maximum value with a period of $0.25$. For this reason, $t\!=\!0.75$ is a carefully selected time where the two-qubit state is well confined in the DFS. Therefore, after the protection is turned off, entanglement remains at its maximum value for a relatively short time, during which the qubits are still in a DFS, and then it starts decreasing as the \textsc{CE} transforms into \textsc{IE} gradually, making DFS disappear. Such a behavior in the dynamics of entanglement can be understood by noting that the first of the aforementioned conditions to form a DFS, namely interaction with a common bath, is no longer satisfied. The results for the most ideal case are presented in Fig.~\ref{dephas}(d) in which case we both turn the protection off and stop altering the position of the qubits after the time $t\!=\!0.75$. In this scenario, one can observe that even if there exists no external field to protect the entanglement between the qubits against the dephasing environment, entanglement remains intact since we keep the qubits in a DFS by maintaining a \textsc{CE}. Therefore, under these conditions, it is actually not necessary to provide an external protection to the qubits at all since the nature of the dynamics guides the system to stay in a DFS. Obviously, the choice of the initial state $\rho_0$ is also relevant here because an initial state defined outside of DFS would not yield the same results. On the other hand, if one intends to keep the protection field on at all times, then the protection field should be sufficiently strong to preserve the entanglement in the system. If the protection the field is not strong enough, i.e. $n_x\!=\!1, 2$, the situation becomes worse than the case of no protection at all since such a case neither provides sufficient protection to reach the levels of entanglement obtained in $n_x\!=\!0$ case, nor lets the system stay in the DFS. All the same, the $n_x\!=\!8$ case in Fig.~\ref{dephas}(d) is an example of how one can use different techniques of preserving the quantumness of a state complementary to each other. Although we have seen that it is not necessary in the present scenario, it should be possible to protect a system from noise by applying a CDD scheme up until the system enters a DFS before turning off the external field. In the next section, we will present an example along these lines.

\subsection{Dephasing and amplitude damping}

\begin{figure*}[t]
\centering
\includegraphics[width=.8\textwidth]{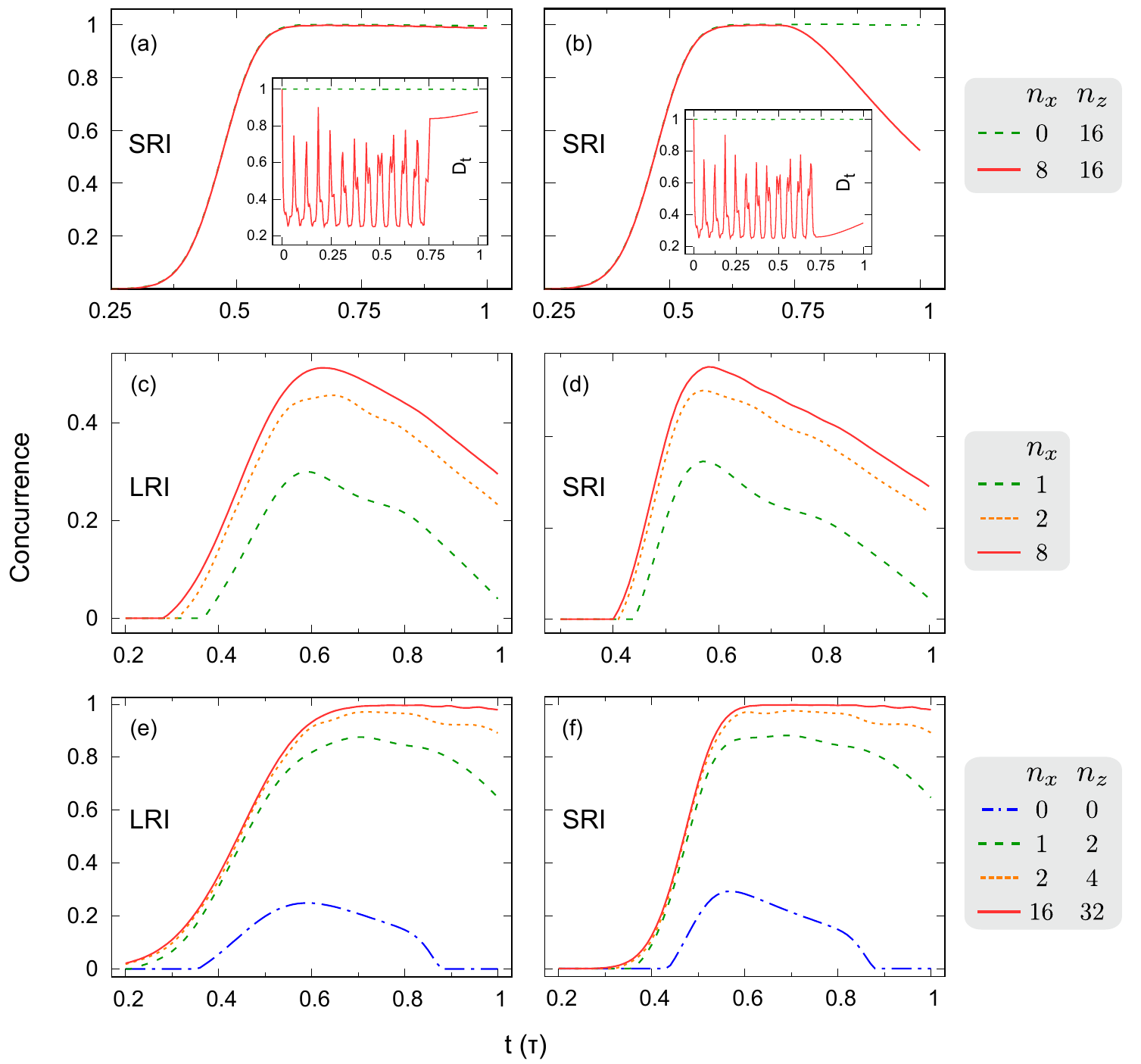}
\caption{Dephasing and amplitude damping affecting the system together: The upper panel, where qubits are halted at (a) $t=0.755$ and (b) $t=0.72$, represents the advantage of leaving dephasing as a residual error, i.e., one can utilize both a DFS and a CDD at the same time to eliminate the undesired effects of the environmental noise. In (a) and (b) the dashed lines represent the case where dephasing is kept as a residual error. For the solid lines, the dephasing protection is applied ($n_x=8$) until the qubits are halted, and then, it is turned off. In the middle panel, marked by (c) and (d), amplitude damping is left as a residual error by setting $n_z=0$, while in the bottom panel, shown by (e) and (f), qubits are protected against the both error mechanisms. No considerable change is observed for larger values of $n_x$ and/or $n_z$.}
\label{resad}
\end{figure*}

In this section, we present our results when both of amplitude damping and dephasing errors are present. We consider two different scenarios for the protection of the qubits' dynamics. On one hand, we provide protection against both decoherence mechanisms. On the other hand, we let either dephasing or amplitude damping affect the system, i.e., supply no protection against it, while providing protection against the other. We also refer to this second case as leaving one of the error mechanisms as a residual error on the dynamics. These studies are relevant since it is experimentally simpler to implement a partial protection compared to the full protection. For example, a simple static field is enough to protect the system, while the full protection requires a more complicated field.

Motivated by the results in Fig.~\ref{dephas} (d), which shows the natural evolution of qubits in a DFS during the dynamics without any protection against dephasing, we want to further investigate this case in a slightly different setting. On top of dephasing, we now assume amplitude damping environment is also acting on our system and we provide full protection against it at all times. Fig.~\ref{resad} (a) and (b) present our results on the described setting and compares the cases of no protection with protection turned off at $t\!=\!0.755$ and $t\!=\!0.72$ for the dephasing, respectively. We know that the entangling gate operation has already ended at these time instants, and we stopped both qubits thereafter to leave them in contact with the CE similar to the case of Fig.~\ref{dephas} (d). Intuitively, one can regard this specific time assignment as a way to gauge how an imperfect timing of shutting the dephasing protective field down would change the entanglement dynamics after that point in time. We observe that in the present setting, it is possible to reach the desired maximally entangled state. The dynamics of the two-qubit density matrix entirely stays inside the subspace spanned by $\mathcal{D}$, which is the same subspace as the DFS occurs. However, it is not possible to directly say that a DFS forms in the present case, due to the fact that there is also an amplitude damping environment and the external field to cancel its effect on the system. Even so, we think the that the perfect generation of the entangled state is because of the reminiscent effect of the DFS. All in all, this result sets an example of the aforementioned hybrid utilization of external and natural protection mechanisms. We now turn our attention to the partial dephasing protection cases that are presented as red solid curves in Fig.~\ref{resad} (a) and (b). Recall that the external field that protects against dephasing errors also makes the state of our system to oscillate in and out of the DFS. We observe that higher confinement inside the DFS at the time we close the dephasing protection results in a slower decay in the entanglement. Since for $n_x\!=\!0$ there is no external dephasing protection field to perturb the system out of the DFS, the performance of the protocol is better in that case as compared to the partial protection. 

In Fig.~\ref{resad}(c) and (d), the amplitude damping mechanism is left as a residual error, that is, there is no protection against it $n_z\!=\!0$. The former is when the interaction is long ranged, while the latter represents the case when the interaction is short ranged. Both figures show qualitatively the same behavior for which in none of them is it possible to reach maximal entanglement. The amount of entanglement follows a decreasing trend after $t\approx 0.5$ after the initial increase. We can conclude that it is possible to reach a certain level of entanglement while leaving amplitude damping channel affecting the qubit system; however, the generated level of entanglement is not sustainable over the course of the dynamics.

In Fig.~\ref{resad}(e) and (f), the behavior of the entanglement is considered for LRI and SRI schemes, respectively. In these figures, we provide protection against both amplitude damping and dephasing for different protection strengths, where $n_x\!=\!n_z\!=\!0$ corresponds to no protection at all for comparison. Examining graphs closely, similar to the previous section, we can conclude that as the external field strength is increased, the CDD scheme works better on the system. Although it is possible to reach the maximal entanglement after the gate operation, the protection is not as stable as the sole dephasing setting even for the highest provided external field strengths $n_x\!=\!16$ and $n_z\!=\!32$. This instability, however, can be further improved by increasing the strength of the CDD field, but we chose to stick to these modest strengths since they are sufficient to demonstrate that proposed protection scenario works well under these circumstances. One more observation is that there is no difference between LRI and SRI cases for any protection strength except that entanglement reaches its maximum slightly earlier in the latter one. This is actually expected since SRI is completed faster than the LRI by definition. 

\section{Conclusion} \label{conclusion}

We proposed a CDD strategy for protecting the dynamics of two qubits from the decohering effects of the environment (amplitude damping and/or dephasing), where we can tune both the inter-qubit interaction strength and qubit--environment couplings. We chose the effective distance between the qubits as our tuning parameter, such that it both adjusts the strength of qubit--qubit interaction and varies the qubit--environment couplings gradually between IE and CE. The qubits get entangled due to the specific type of interaction between them, and we employed concurrence as the figure of merit for our protection scheme. We showed that the entangling gate operation, which is mediated by the Heisenberg interaction, can be preserved almost perfectly if the strength of the external protective field is strong enough for both decoherence channels. Although the presented model has its own limitations and relies on a set of assumptions, we believe that it could potentially serve as an example to provide a direction for the protection of entanglement in open quantum systems.

An interesting finding in this work is the possibility of utilizing DFS in the Hilbert space of the qubits under some specific conditions. We observed that if the inter-qubit coupling starts and ends inside the region where the qubits are in contact only with a common dephasing environment, the natural occurrence of DFS due to the CE dephasing dynamics perfectly preserves the entangled state, even in the absence of any external protection mechanism. The very presence of a CDD field actually destroys the DFS, however, what one can do is to turn this external protection off once the system enters into the DFS, guaranteeing that it remains unaffected from the environmental noise afterward. We demonstrated that provided that the field cannot be turned off completely, then its strength should be over a certain threshold value; otherwise, the external field may drive the state of the system out of the DFS and makes the situation worse than the case of no external field. Furthermore, we showed that when amplitude damping errors are also present, it is sufficient to provide external protection only for amplitude damping while the system is protected from dephasing naturally by evolving into a DFS-like subspace. We hope that the strategy presented in our example model might be of interest as it combines two of the well-known methods to prevent the undesired effects of an environment from destroying quantum features of a system. Besides, it partially relies on external resources (CDD) up to a point where it can use the internal mechanisms arising from the nature of the dynamics (DFS).

\appendix*

\section{Derivation of the master equation}

We introduce the details of the derivation of the master equation that governs the time evolution of our two-qubit system. First of all, we assume that each source of error, induced by the environment, is present separately. In other words, both independent environments for each of the qubits and the common environment, which the qubits together couple when they are close, are assumed to be present during the whole evolution. In this case, the total Hamiltonian is given by Eq. (\ref{dd}), with $H_{\rm int}$ defined by Eq.~(\ref{eq:gradChEnv}) and $H_{\rm env}$ defined by:
\begin{equation}
H_{\rm env}=\sum _{k} \omega_{k} {a_{k}}^{\dagger}a_{k}+\sum_{s=1}^{2}\sum_{k} \omega_{k}^{(s)}{a_{k}^{(s)}}^{\dagger}a_{k}^{(s)}, 
\end{equation}
where $\omega_{k}$ is the frequency of the $k$th normal mode of the environment that introduces common errors, with $a_{k}$ and ${a_{k}}^\dagger$ being the annihilation and creation operators, respectively. In the second term, $\omega _{k}^{(s)}$ is the frequency of the $k$th normal mode of the $s$th independent environment with respective annihilation ($a_{k}^{(s)}$) and creation (${a_{k}^{(s)}}^{\dagger}$) operators. It is important to emphasize that each error source is present independently of  others since, in such a case, the total environment Hamiltonian could be written as a linear combination of the Hamiltonians of three distinct environments.

Having defined the total Hamiltonian, we utilize the Redfield master equation approach to calculate time evolution of the reduced two-qubit system:
\begin{equation}
\frac{d\tilde{\rho}_{S}\left(t\right)}{dt} =  -\int_{0}^{t}\Tr_{\rm env}\left\{ \left[\tilde{H}_{\rm int}\left(t\right),\left[\tilde{H}_{\rm int}\left(t^{\prime}\right),\rho_{R}\tilde{\rho}_{S}\left(t\right)\right]\right]\right\} dt^{\prime}.
\end{equation}
Here, 
\begin{equation}
\tilde{H}_{\rm int}(t)=U_{\rm env}^\dagger(t) U_{\rm gate}^\dagger(t) U_{c}^\dagger(t) {H}_{\rm int}U_{c}(t) U_{\rm gate}(t) U_{\rm env}(t),
\end{equation}
where $U_{\rm env}(t) = \exp(-iH_{\rm env}t)$, $U_{\rm gate}(t)$ is the unitary evolution operator related to the time-dependent Hamiltonian $H_{\rm gate}(t)$, given by Eq. (\ref{H0}), and $U_c(t)$ is given by Eq. (\ref{Uc}). In addition, $\rho_E=\frac{1}{Z}\exp(-\beta H_{\rm env})$, where $Z$ is the partition function $Z = {\rm Tr_{env} }\exp(-\beta H_{\rm env})$ and $\beta=1/k_BT$ with $k_B$ being the Boltzmann constant, and $T$ is the absolute temperature of the environment. Finally, $\tilde{\rho}_{S}\left(t\right) = U_{\rm gate}^\dagger \left(t\right) U_{c}^\dagger \left(t\right) {\rho}_{S}\left(t\right)U_{c} \left(t\right) U_{\rm gate} \left(t\right) $ where ${\rho}_{S}(t)$ is the density operator in the Schrodinger picture.

To write an effective master equation in order to be solved numerically, we rewrite the interaction Hamiltonian as 
\begin{equation}
H_{\rm int}=\sum_{n=1}^3 \bm {\lambda}\cdot {\bm S}_n(t)\otimes B_n+\bm{\lambda^\ast} \cdot {\bm S}_n^{\dagger}(t)
\otimes B_n^{\dagger},
\end{equation}
where $n=\{1,2,3\}$ represents each one of the three independent baths (one individual for each qubit plus a collective one for both qubits) and 
\begin{eqnarray}
{\bm S}_1(t) &=& \xi(t)\left[{\bm \sigma}^{(1)} + {\bm \sigma}^{(2)}\right],\nonumber\\
{\bm S}_2(t) &=& \left[1-\xi(t)\right]{\bm \sigma}^{(1)},\nonumber\\
{\bm S}_3(t) &=& \left[1-\xi(t)\right]{\bm \sigma}^{(2)},\\\nonumber
\end{eqnarray}
with $B_n = \sum_k g_k^{n}a_k^{n}$, and ${\bm \lambda} = \hat z$ for a dephasing environment and ${\bm \lambda}=(\hat x + i \hat y)$ for the amplitude damping. In the interaction picture, we can finally write the interaction Hamiltonian as:
\begin{equation}
\tilde{H}_{\rm int}\left(t\right)  =  \sum_{n=1}^3 \Lambda_n\left(t\right)\tilde{B}_n(t)+\Lambda^{\ast}_n\left(t\right)\tilde{B}_n^{\dagger}\left(t\right),
\end{equation}
where 
\begin{eqnarray}
\Lambda_n\left(t\right)&=&U_{\rm gate}^\dagger \left(t\right) U_{c}^\dagger \left(t\right)\left[{\bm \lambda \cdot {\bm S} }_n(t)\right]U_{c}\left(t\right)U_{\rm gate}\left(t\right),\\
\tilde{B}_n(t)&=&U_{\rm env}^{\dagger}\left(t\right) (B_n) U_{\rm env}\left(t\right).
\end{eqnarray}
Replacing $\tilde{H}_{\rm int}\left(t\right)$ in the master equation, we can write it in a more clear way:
\begin{widetext}
\begin{eqnarray}
\frac{d\tilde{\rho}_{S}\left(t\right)}{dt} =&\sum_{n=1}^3 \left[\tilde{\rho}_{S}\left(t\right)\Lambda_n^{\ast}\left(t^{\prime}\right),\Lambda_n\left(t\right)\right]{\cal G}_{1}\left(t-t^{\prime}\right)+\left[\tilde{\rho}_{S}\left(t\right)\Lambda_n\left(t^{\prime}\right),\Lambda_n^{\ast}\left(t\right)\right]{\cal G}_{2}\left(t-t^{\prime}\right)\nonumber\\+&\left[\Lambda_n^{\ast}\left(t\right),\Lambda_n\left(t^{\prime}\right)\tilde{\rho}_{S}\left(t\right)\right]{\cal G}_{1}^{\ast}\left(t-t^{\prime}\right)+\left[\Lambda_n\left(t\right),\Lambda_n^{\ast}\left(t^{\prime}\right)\tilde{\rho}_{S}\left(t\right)\right]{\cal G}_{2}^{\ast}\left(t-t^{\prime}\right),
\end{eqnarray}
\end{widetext}
where 
\begin{eqnarray}
{\mathcal G}_1(t) &=& \int_0^\infty d\omega J(\omega)h(\omega)\exp(-i\omega t)\nonumber\\
{\mathcal G}_2(t) &=& \int_0^\infty d\omega J(\omega)\exp(i\omega t)[h(\omega)+1],\nonumber\\
\end{eqnarray}
with $h(\omega) = \frac{1}{\exp(\beta \omega) -1}$, and $J(\omega) =  \omega \exp(-\omega/\omega_c)$ where $\omega_c$ is the cutoff frequency. Note that to describe the environment spectrum, as usual, we assume that the number of environmental normal modes per unit frequency becomes very large. We also assume that all environments are identical since ${\mathcal G}_1(t)$ and ${\mathcal G}_2(t)$ are same for all baths. Further details can be found in Ref.~\cite{fanchini2015}, where the calculation has been developed for the case of time independent interaction Hamiltonians.

\begin{acknowledgements}
\.{I}.Y.\ is supported by the Project RVO 68407700 and RVO 14000 and funding from the project “Centre for Advanced Applied Sciences,” Registry No. CZ.02.1.01/0.0/0.0/16 019/0000778, supported by the Operational Programme Research, Development and Education, co-financed by the European Structural and Investment Funds and the state budget of the Czech Republic. F.F.F. has been supported by the Brazilian agencies FAPESP under Grant No. 2017/07787-7, by CNPq under Grant No. 302280/2017-0, and by INCT-IQ. G.K. is supported by the BAGEP Award of the Science Academy and the GEBIP program of the Turkish Academy of Sciences (TUBA). G.K. is also supported by the Scientific and Technological Research Council of Turkey (TUBITAK) under Grant No. 117F317.
\end{acknowledgements}

\bibliography{bibliography}

\end{document}